\documentclass[aps,pra,twocolumn,nofootinbib]{revtex4}
\usepackage{amsmath, amsthm, amscd, amssymb}
\usepackage{bm}
\usepackage{bbm}
\usepackage{epsfig}

\begin{document}

\title{Radiation from Atoms Falling into a Black Hole }
\author{Marlan O. Scully$^{1,2,3}$, Stephen Fulling$^1$, David Lee$^1$, Don
Page$^4$, Wolfgang Schleich$^{1,5}$, Anatoly Svidzinsky$^1$}
\affiliation{$^1$Texas A\&M University, College Station, TX 77843; \\
$^2$Baylor University, Waco, TX 76798; \\
$^3$Princeton University, Princeton, NJ 08544; \\
$^4$University of Alberta, Edmonton, Canada T6G 2E1; \\
$^5$Universit\"at Ulm, Germany D-89081 }
\date{\today }

\begin{abstract}
We show that atoms falling from outside through a cavity into a black hole
(BH) emit acceleration radiation which to a distant observer looks much like
Hawking BH radiation. In particular, we find the entropy of the acceleration
radiation via a simple laser-like analysis. We call this entropy Horizon
Brightened Acceleration Radiation (HBAR) entropy to distinguish it from the
BH entropy of Bekenstein and Hawking.
\end{abstract}

\maketitle

\section{Introduction}

General relativity as originally developed by Einstein \cite{Eins15} is
based on the union of geometry and gravity \cite{Misn73}. Half a century
later the union of general relativity and thermodynamics was found to yield
surprising results such as Bekenstein-Hawking black hole entropy \cite%
{Beke73,Hawk74}, particle emission from a black hole \cite{Hawk74,Page76}
and acceleration radiation \cite{Unru76}. More recently the connection
between black hole (BH) physics and optics, e.g., ultraslow light \cite%
{Weis00}, fiber-optical analog of the event horizon \cite{Phil08} and
quantum entanglement \cite{Das06} has led to fascinating physics.

In their seminal works, Hawking, Unruh and others showed how quantum effects
in curved space yield a blend of thermodynamics, quantum field theory and
gravity which continues to intrigue and stimulate. For problems as important
and startling as Hawking and Unruh radiation, new and alternative approaches
are of interest. In that regard it was shown \cite{Scul03} that virtual
processes in which atoms jump to an excited state while emitting a photon is
an alternative way to view Unruh acceleration radiation. Namely, by breaking
and interrupting the virtual processes which take place all around us we can
render the virtual photons real.

\begin{figure}[t]
\begin{center}
\epsfig{figure=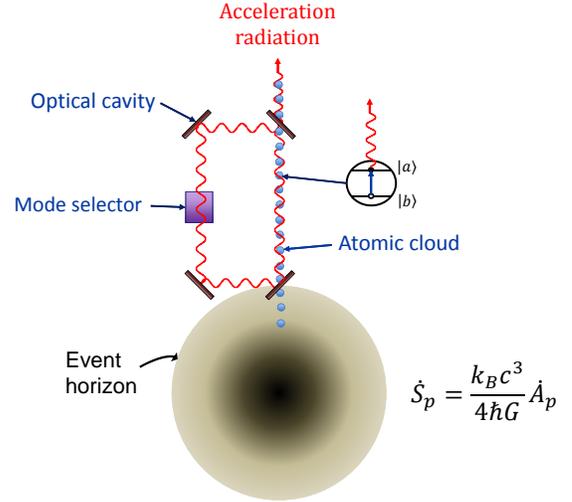, angle=270, width=8.5cm}
\end{center}
\caption{A BH is bombarded by a pencil-like cloud of two level atoms falling
radially from infinity. A cavity is held at the event horizon which shields
infalling atoms from the Hawking radiation and the mode selector picks one
cavity mode (or a few modes) counterpropagating relative to the atoms. The
relative acceleration between the atoms and the field yields generation of
acceleration radiation. The physics of the acceleration radiation process
corresponds to the excitation of the atom together with the emission of the
photon (see Appendix \protect\ref{A2}).}
\label{Fig1}
\end{figure}

The present paper is an extension of that logic by considering what happens
when atoms fall through the Boulware vacuum \cite{Bulw75} into a BH as shown
in Fig. \ref{Fig1}. The equivalence principle tells us that an atom falling
in a gravitational field does not \textquotedblleft feel\textquotedblright\
the effect of gravity, namely its $4-$acceleration is equal to zero.
However, as we discuss in Appendix \ref{A1}, there is relative acceleration
between the atoms and the field modes. This leads to the generation of
acceleration radiation. In Appendix \ref{A2} we provide a detailed
calculation of the photon emission by atoms falling into a BH.

Specifically we consider an atomic cloud consisting of two level atoms
emitting acceleration radiation (see Fig. \ref{Fig1}) \cite{Scul03}. We find
that the quantum master equation technique, as developed in the quantum
theory of the laser \cite{Scul66}, provides a useful tool for the analysis
of BH acceleration radiation and the associated entropy. In particular, we
derive a coarse grained equation of motion for the density matrix of the
emitted radiation of the form

\begin{equation}
\dot{\rho}_{nn}=\left( \mathcal{M}\rho \right) _{nn},
\end{equation}%
where the time evolution of the diagonal elements of the density matrix $%
\rho _{nn}$ is governed by the super operator $\mathcal{M}$ as given by Eq. (%
\ref{R1}).

Furthermore, we find that once we have cast the acceleration radiation
problem in the language of quantum optics and cavity QED the entropy follows
directly. Specifically, once we calculate $\dot{\rho}$ for the field
produced by accelerating atoms, we can use the von Neumann entropy to write 
\begin{equation}
\dot{S}_{p}=-k_{B}\text{Tr}(\dot{\rho}\ln \rho )  \label{h2}
\end{equation}%
to calculate the radiation entropy flux directly. From the present
perspective the acceleration radiation - BH entropy problem is close in
spirit to the quantum theory of the laser.

Hawking's pioneering proof that BHs are not black \cite{Hawk74} is based on
a quantum field theoretic analysis showing that photon emission from a BH is
characterized by a temperature $T_{\text{BH}}$ and generalized BH entropy.
James York \cite{York83} gives an analogy between radiation from a BH and
total internal reflection in classical optics. He argues that a light beam
in a dense medium can undergo total internal reflection at a flat optical
surface; but if we sprinkle dust particles on the surface some light will be
transmitted. Now the flat surface can be likened to the BH event horizon,
the dust is replaced by vacuum fluctuations and light is transmitted through
the horizon.

Hawking showed that the radiation that comes out from the BH\ is described
by the temperature 
\begin{equation}
T_{\text{BH}}=\frac{\hbar c^{3}}{8\pi k_{B}GM}.  \label{TBH}
\end{equation}%
Hawking then associates the energy of the emitted radiation $\delta E$ with
the loss in energy of the BH $\delta (Mc^{2})$ and writes the entropy loss
as $\delta S=\delta (Mc^{2})/T_{\text{BH}}$. Then using Eq. (\ref{TBH}) he
obtains%
\begin{equation}
\delta S=k_{B}\frac{8\pi G}{\hbar c}M\delta M=\frac{k_{B}c^{3}}{4\hbar G}%
\delta A,  \label{SBH}
\end{equation}%
where the BH area in terms of the gravitational radius $r_{g}=2GM/c^{2}$ is
given by $A\equiv 4\pi r_{g}^{2}=16\pi G^{2}M^{2}/c^{4}$.

In the present paper we analyze the problem of atoms outside the event
horizon emitting acceleration radiation as they fall into the BH. The
emitted radiation is essentially, but not precisely, thermal and has an
entropy analogous to the BH result given by Eq. (\ref{SBH}). However, the
physics is very different. Here we have radiation coming from the atoms,
whereas Hawking radiation requires no extra matter (e.g. atoms). Indeed,
Hawking radiation arises just from the BH geometry and the initial state of
the quantized field involved.

Historically, Bekenstein \cite{Beke73} introduced the BH entropy concept by
information theory arguments. Hawking \cite{Hawk74} then introduced the BH
temperature to calculate the entropy. In the present approach we calculate
the radiation density matrix and then calculate the entropy directly. To
distinguish this from the BH entropy we call it the Horizon Brightened
Acceleration Radiation (HBAR) entropy.

\section{The HBAR entropy via quantum statistical mechanics}

As noted earlier, we here consider a BH bombarded by a beam of two-level
atoms with transition frequency $\omega $ which fall into the event horizon
at a rate $r$ (see Fig. \ref{Fig1}). The atoms emit and absorb the
acceleration radiation.

We seek the density matrix of the field. As in the quantum theory of the
laser \cite{Scul66}, the (microscopic) change in the density matrix of the
field due to any one atom, $\delta \rho ^{i}$, is small. The (macroscopic)
change due to $\Delta N$ atoms is then%
\begin{equation}
\Delta \rho =\sum_{i}\delta \rho ^{i}=\Delta N\delta \rho .
\end{equation}%
Writing $\Delta N=r\Delta t$, where $r$ is the atomic injection rate, we
have the coarse grained equation of motion%
\begin{equation}
\frac{\Delta \rho }{\Delta t}=r\delta \rho .
\end{equation}%
We thus obtain an evolution equation for the radiation following the
approach used in the quantum theory of the laser \cite{Scul66}. As is
further discussed in Appendices \ref{A2} and \ref{A3}, the coarse grained
time rate of change of the radiation field density matrix for a particular
field mode is found to be%
\begin{equation*}
\frac{1}{R}\frac{d\rho _{n,n}}{dt}=-\frac{rg^{2}}{\omega ^{2}}e^{-\xi }\left[
(n+1)\rho _{n,n}-n\rho _{n-1,n-1}\right]
\end{equation*}%
\begin{equation}
-\frac{rg^{2}}{\omega ^{2}}e^{\xi }\left[ n\rho _{nn}-(n+1)\rho _{n+1,n+1}%
\right] ,  \label{R1}
\end{equation}%
where $g$ is the atom-field coupling constant, $\xi =2\pi \nu r_{g}/c$, 
\begin{equation}
R=\frac{\xi }{\sinh (\xi )}
\end{equation}%
and $\nu $ is the photon frequency far from the BH. Using Eqs. (\ref{h2})
and (\ref{R1}), we find that the von-Neumann entropy generation rate of the
HBAR is (see Appendix \ref{A4} for details)%
\begin{equation}
\dot{S}_{p}=\frac{4\pi k_{B}r_{g}}{c}\sum_{\nu }\dot{\bar{n}}_{\nu }\nu \,,
\label{s4}
\end{equation}%
where $\dot{\bar{n}}_{\nu }$ is the flux of photons with frequency $\nu $
coming from the cavity and propagating away from the BH.

Taking into account that the BH mass change due to photon emission is $\dot{m%
}_{p}c^{2}=\hbar \sum_{\nu }\dot{\bar{n}}_{\nu }\nu $, we arrive at the HBAR
entropy/area relation%
\begin{equation}
\dot{S}_{p}=\frac{k_{B}c^{3}}{4\hbar G}\dot{A}_{p}\,.  \label{Sdot}
\end{equation}%
Here $\dot{A}_{p}=$ $(2\dot{m}_{p}/M)A$ is the rate of change of the BH area
due to photon emission which we are interested in; see Appendix \ref{A4}.

\section{Discussion and Summary}

Conversion of virtual photons into directly observable real photons is a
subject not without precedent. Moore's accelerating mirrors \cite{Moor70},
the rapid change of refractive index considered by Yablonovitch \cite{Yabl89}
and the more recent observation of the Dynamical Casimir effect in a
superconducting circuit \cite{Wils11} are a few examples.

The physics behind acceleration radiation is explained in Ref.~\cite{Scul03}
(see also \cite{Marz98}) where it is stated that:

\begin{quote}
In conclusion our simple model demonstrates that the ground-state atoms
accelerated through a field vacuum-state radiate real photons. ... The
physical origin of the field energy in the cavity and of the internal energy
in the atom is the work done by an external force driving the center-of-mass
motion of the atom against the radiation reaction force. Both the present
single mode and the many mode effect originate from the transition of the
ground-state atom to the excited state with simultaneous emission of photon
due to the counter-rotating terms in the Hamiltonian.
\end{quote}

In other words the virtual processes in which an atom jumps from the ground
state to an excited state, together with the emission of a photon, followed
by the reabsorption of the photon and return to the ground state, are
altered by the acceleration. The atom is accelerated away from the original
point of virtual emission, and there is a small probability that the virtual
photon will \textquotedblleft get away\textquotedblright\ before it is
reabsorbed\ as is depicted in Fig. \ref{Fig1}.

\begin{figure}[t]
\begin{center}
\epsfig{figure=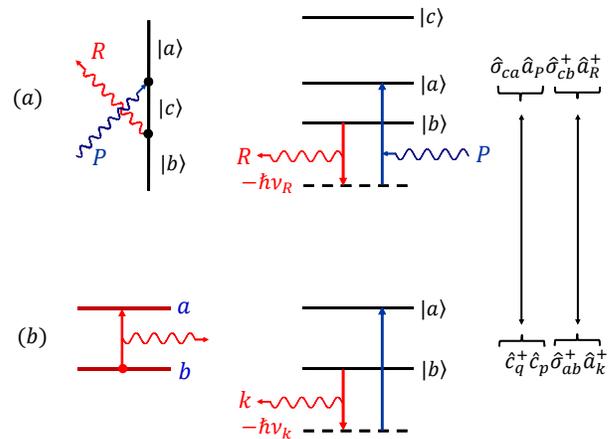, angle=270, width=8.5cm}
\end{center}
\caption{(a) Processes involved in Raman scattering in which emission occurs
before absorption. A molecule promoted from $\left\vert b\right\rangle $ to $%
\left\vert c\right\rangle $ (virtually) with emission of a Raman photon then
absorbs a pump photon while the molecule drops to a state $\left\vert
a\right\rangle $. Such a process is due to counter-rotating terms in the
Hamiltonian much as in the case of acceleration radiation. (b) Processes
involved in acceleration radiation of a two-level atom. Operators $\hat{c}%
_{p}$ and $\hat{c}_{q}^{+}$ describe the change in the center of mass
motion. An analogy between elements of Raman and acceleration radiation
processes are shown at the extreme right. }
\label{Raman}
\end{figure}

The Raman effect provides another precedent for the acceleration radiation
problem. There are two types of processes taking place in ordinary Raman
scattering. (1) The higher frequency pump is absorbed followed by emission
of a lower frequency Raman photon. (2) The other process (Fig. \ref{Raman}a)
involves the molecule going into a virtual state and at the same time
emitting a Raman photon, then a pump photon is absorbed. The excitation of
the molecule and emission of photons is thus said to take place before
absorption. A similar process occurs in the acceleration radiation and
involves, instead of a pump field, a change in the center of mass motion
governed by the operators $\hat{c}_{p}$ and $\hat{c}_{q}^{+}$ (Fig. \ref%
{Raman}b).

Acceleration radiation involves a combination of two effects: acceleration
and nonadiabaticity that produce the emitted light. The energy is supplied
by the external force field (e.g., the gravitational field of the star).

Gravitational acceleration of atoms is also a source of confusion. The
equivalence principle tells us that the atom essentially falls
\textquotedblleft force free\textquotedblright\ into the BH. How can it then
be radiating? Indeed, the atomic evolution in the atom frame is described by
the $e^{i\omega \tau }$ term in the Hamiltonian (\ref{u2}). From the
Hamiltonian we clearly see that it is the photon time (and space) evolution
which contain effective acceleration. The radiation modes are fixed relative
to the distant stars, and the photons (not the atoms) carry the seed of the
acceleration effects in $\hat{V}(\tau )$. The fact that a freely falling
atom (detector) is excited and emits radiation is nicely explained in \cite%
{Ginz87,Ahma14}.

The present model is simple enough to allow a direct calculation of the HBAR
entropy. It is a much more tractable problem then the daunting BH entropy
issue. It is interesting that the answer for the HBAR entropy we found is
essentially the same as the formula for the Bekenstein-Hawking black hole
entropy.

Perhaps the quantum master equation approach can provide a useful tool for
calculating the latter. That is, by regarding the material inside the
horizon as a reservoir (in some sense like the atoms in the present
approach), one can perhaps derive an equation of motion for the density
matrix of the Hawking radiation along the lines of Eqs. (\ref{B1}) and (\ref%
{u3}) and then calculate the entropy of the Hawking radiation. We plan to
address these and other related questions elsewhere.

We wish to thank M. Becker, S. Braunstein, C. Caves, G. Cleaver, S. Deser,
E. Martin-Martinez, G. Moore, W. Unruh and A. Wang for helpful discussions.
This work was supported by the Office of Naval Research (Award Nos.
N00014-16-1-3054 and N00014-16-1-2578), the National Science Foundation (Award
No. DMR 1707565), the Robert A. Welch Foundation (Award A-1261), and the
Natural Sciences and Engineering Research Council of Canada.

\appendix

\section{Motion of particle in Rindler and Schwarzschild space-time}

\label{A1}

When atoms are in free fall their operator time dependence in the
interaction picture goes as $\hat{\sigma}^{+}(\tau )=\hat{\sigma}%
^{+}(0)e^{i\omega \tau }$, where $\tau $ is the proper time of the atom. The
corresponding time evolution of the radiation field operator is $\hat{a}%
_{k}^{+}(t)=\hat{a}_{k}^{+}(0)\psi \left[ t(\tau ),z(\tau )\right] $, where $%
\psi (t,z)$ is the mode function and the space and time parametrization of
the field $t(\tau )$ and $z(\tau )$ are to be determined. In what follows we
obtain the results in three steps: (1) Special relativity, (2) Rindler
metric, (3) Schwarzschild metric.

\subsection{Special Relativity}

First of all we note that finding $t(\tau )$ and $z(\tau )$ i.e. the
coordinate time and position of the atom in terms of the atom's proper time
is really a problem in special relativity. Namely, from the $2D$ Minkowski
line element 
\begin{equation}
ds^{2}=c^{2}dt^{2}-dz^{2}  \label{min}
\end{equation}%
we can write 
\begin{equation}
\tau =\int_{0}^{\tau }d\tau =\int_{0}^{t}\sqrt{1-\frac{V^{2}}{c^{2}}}dt,
\end{equation}%
where $V=dz/dt$. One can show that for a particle moving with constant
proper acceleration $a$ 
\begin{equation}
V=\frac{at}{\sqrt{1+\frac{a^{2}t^{2}}{c^{2}}}}
\end{equation}%
and, therefore, 
\begin{equation}
\tau =\int_{0}^{t}\frac{dt}{\sqrt{1+\frac{a^{2}t^{2}}{c^{2}}}}=\frac{c}{a}%
\sinh ^{-1}\left( \frac{at}{c}\right) ,  \label{tau}
\end{equation}%
or%
\begin{equation}
t(\tau )=\frac{c}{a}\sinh \left( \frac{a\tau }{c}\right) .  \label{a1}
\end{equation}%
Likewise, integration of $V(t)$ yields%
\begin{equation}
z(t)-z(0)=\int_{0}^{t}V(t)dt=\frac{c^{2}}{a}\left( \sqrt{1+\frac{a^{2}t^{2}}{%
c^{2}}}-1\right) .
\end{equation}%
Setting $z(0)=c^{2}/a$ and using Eq. (\ref{a1}) we obtain%
\begin{equation}
z(\tau )=\frac{c^{2}}{a}\cosh \left( \frac{a\tau }{c}\right) .  \label{a2}
\end{equation}

\subsection{Rindler}

The Rindler metric for a particle undergoing uniformly accelerated motion is
obtained from the Minkowski line element (\ref{min}) if we make a coordinate
transformation 
\begin{equation}
t=\frac{\bar{z}}{c}\sinh \left( \frac{\bar{a}\bar{t}}{c}\right) ,  \label{z1}
\end{equation}%
\begin{equation}
z=\bar{z}\cosh \left( \frac{\bar{a}\bar{t}}{c}\right) ,  \label{z2}
\end{equation}%
where $\bar{a}$ is a constant. This leads to the line element%
\begin{equation}
ds^{2}=\left( \frac{\bar{a}\bar{z}}{c^{2}}\right) ^{2}c^{2}d\bar{t}^{2}-d%
\bar{z}^{2},
\end{equation}%
which is the Rindler line element describing uniformly accelerated motion.
Comparison of Eqs. (\ref{z1}) and (\ref{z2}) with Eqs. (\ref{a1}) and (\ref%
{a2}) shows that a particle moving along a trajectory with constant $\bar{z}$
in Rindler space has $\tau =\bar{a}\bar{t}/a$ and is uniformly accelerating
in Minkowski space with acceleration 
\begin{equation}
a=\frac{c^{2}}{\bar{z}}.  \label{z3}
\end{equation}

\subsection{Schwarzschild}

Finally we make an observation that the $t-r$ part of the Schwarzschild
metric,%
\begin{equation}
ds^{2}=\left( 1-\frac{r_{g}}{\bar{r}}\right) c^{2}d\bar{t}^{2}-\frac{1}{1-%
\frac{r_{g}}{\bar{r}}}d\bar{r}^{2},  \label{gs1}
\end{equation}%
where $r_{g}=2GM/c^{2}$ is the gravitational radius, can be approximated
around $r_{g}$ by Rindler space by using the coordinate $0<\bar{z}\ll r_{g}$
defined by%
\begin{equation}
\bar{r}=r_{g}+\frac{\bar{z}^{2}}{4r_{g}}.
\end{equation}%
Expanding around $r_{g}$%
\begin{equation}
1-\frac{r_{g}}{\bar{r}}\approx \frac{\bar{z}^{2}}{4r_{g}^{2}}
\end{equation}%
yields the Rindler metric \cite{Rind60} 
\begin{equation}
ds^{2}=\frac{\bar{z}^{2}}{4r_{g}^{2}}c^{2}d\bar{t}^{2}-d\bar{z}^{2}.
\end{equation}%
According to Eq. (\ref{z3}), curves of constant $\bar{z}$ (or $\bar{r}$)
correspond to uniformly accelerated motions with 
\begin{equation}
a=\frac{c^{2}}{\bar{z}}=\frac{c^{2}}{2r_{g}}\frac{1}{\sqrt{1-\frac{r_{g}}{%
\bar{r}}}}.  \label{AC1}
\end{equation}

\section{Acceleration radiation from atoms falling into a black hole}

\label{A2}

Here we consider a two-level ($a$ is the excited level and $b$ is the ground
state) atom with transition angular frequency $\omega $ freely falling into
a nonrotating BH of mass $M$ along a radial trajectory from infinity with
zero initial velocity. We choose the gravitational radius $r_{g}=2GM/c^{2}$
as a unit of distance and $r_{g}/c$ as a unit of time and introduce the
dimensionless distance, time and frequency as%
\begin{equation*}
r\rightarrow r_{g}r,\quad t\rightarrow (r_{g}/c)t,\quad \omega \rightarrow
(c/r_{g})\omega .
\end{equation*}

In dimensionless Schwarzschild coordinates the atom trajectory is described
by the equations 
\begin{equation}
\frac{dr}{d\tau }=-\frac{1}{\sqrt{r}},\quad \frac{dt}{d\tau }=\frac{r}{r-1},
\label{u1}
\end{equation}%
where $t$ is the dimensionless time in Schwarzschild coordinates and $\tau $
is the dimensionless proper time for the atom. Integration of equations (\ref%
{u1}) yields%
\begin{equation}
\tau =-\frac{2}{3}r^{3/2}+const,  \label{m3}
\end{equation}%
\begin{equation}
t=-\frac{2}{3}r^{3/2}-2\sqrt{r}-\ln \left( \frac{\sqrt{r}-1}{\sqrt{r}+1}%
\right) +const.  \label{m4}
\end{equation}%
For a scalar photon in the Regge-Wheeler coordinate%
\begin{equation}
r_{\ast }=r+\ln (r-1)  \label{m5}
\end{equation}%
the field propagation equation reads%
\begin{equation}
\left[ \frac{\partial ^{2}}{\partial t^{2}}-\frac{\partial ^{2}}{\partial
r_{\ast }^{2}}+\left( 1-\frac{1}{r}\right) \left( \frac{1}{r^{3}}-\frac{%
\Delta }{r^{2}}\right) \right] \psi =0,  \label{m1}
\end{equation}%
where $\Delta $ is the angular part of the Laplacian. We are interested in
solutions of this equation outside of the event horizon, that is for $r>1$.
If the dimensionless photon angular frequency $\nu \gg 1$, then the first
two terms in Eq. (\ref{m1}) dominate and one can approximately write 
\begin{equation}
\left( \frac{\partial ^{2}}{\partial t^{2}}-\frac{\partial ^{2}}{\partial
r_{\ast }^{2}}\right) \psi =0.
\end{equation}%
We consider a solution of this equation describing an outgoing wave 
\begin{equation}
\psi =e^{i\nu \left( t-r_{\ast }\right) }=e^{i\nu \left[ t-r-\ln (r-1)\right]
},  \label{m2}
\end{equation}%
where $\nu $ is the wave frequency measured by a distant observer. In
general we will have many modes of the field (frequencies $\nu $) which we
will sum over as in Eq. (\ref{s4}).

The interaction Hamiltonian between the atom and the field mode (\ref{m2})
is 
\begin{equation}
\hat{V}(\tau )=\hbar g\left[ \hat{a}_{\nu }e^{-i\nu t(\tau )+i\nu r_{\ast
}(\tau )}+\text{H.c.}\right] \left( \hat{\sigma}e^{-i\omega \tau }+\text{H.c.%
}\right) ,  \label{u2}
\end{equation}%
where the operator $\hat{a}_{\nu }$ is the photon annihilation operator, $%
\hat{\sigma}$ is the atomic lowering operator and $g$ is the atom-field
coupling constant. We assume that $g\approx $const which is the case for a
scalar (spin-0) \textquotedblleft photons\textquotedblright . Initially the
atom is in the ground state and there are no photons for the modes with
frequency $\nu $, so that the field is in the Boulware vacuum \cite{Bulw75}.

The probability of excitation of the atom (frequency $\omega $) with
simultaneous emission of a photon with frequency $\nu $ is due to a
counterrotating term $\hat{a}_{\nu }^{+}\hat{\sigma}^{+}$ in the interaction
Hamiltonian. The probability of this event,%
\begin{equation*}
P_{exc}=\frac{1}{\hbar ^{2}}\left\vert \int d\tau \left\langle 1_{\nu
},a\right\vert \hat{V}(\tau )\left\vert 0,b\right\rangle \right\vert ^{2}
\end{equation*}%
\begin{equation*}
=g^{2}\left\vert \int d\tau e^{i\nu t(\tau )-i\nu r_{\ast }(\tau
)}e^{i\omega \tau }\right\vert ^{2},
\end{equation*}%
can be written as an integral over the atomic trajectory from $r=\infty $ to
the event horizon $r=1$ as%
\begin{equation}
P_{exc}=g^{2}\left\vert \int_{\infty }^{1}dr\left( \frac{d\tau }{dr}\right)
e^{i\nu t(r)-i\nu r_{\ast }(r)}e^{i\omega \tau (r)}\right\vert ^{2}.
\end{equation}%
Inserting here Eqs. (\ref{m3})-(\ref{m5}) we obtain%
\begin{equation*}
P_{exc}=g^{2}\left\vert \int_{1}^{\infty }dr\sqrt{r}e^{-i\nu \left[ \frac{2}{%
3}r^{3/2}+r+2\sqrt{r}+2\ln \left( \sqrt{r}-1\right) \right] }e^{-\frac{2}{3}%
i\omega r^{3/2}}\right\vert ^{2}.
\end{equation*}%
Making change of the integration variable into $y=r^{3/2}$ yields%
\begin{equation}
P_{exc}=\frac{4g^{2}}{9}\left\vert \int\limits_{1}^{\infty }dye^{-i\nu \left[
\frac{2}{3}y+y^{2/3}+2y^{1/3}+2\ln \left( y^{1/3}-1\right) \right] }e^{-%
\frac{2}{3}i\omega y}\right\vert ^{2}.  \label{u4}
\end{equation}%
Next we make another change of the integration variable $x=\frac{2\omega }{3}%
(y-1)$ and find%
\begin{equation}
P_{exc}=\frac{g^{2}}{\omega ^{2}}\left\vert \int\limits_{0}^{\infty
}dxe^{-i\nu \phi (x)}e^{-ix}\right\vert ^{2},  \label{m6}
\end{equation}%
where%
\begin{equation*}
\phi (x)=\frac{x}{\omega }+\left( 1+\frac{3x}{2\omega }\right)
^{2/3}+2\left( 1+\frac{3x}{2\omega }\right) ^{1/3}
\end{equation*}%
\begin{equation*}
+2\ln \left[ \left( 1+\frac{3x}{2\omega }\right) ^{1/3}-1\right] .
\end{equation*}

The asymptotic behavior of Eq. (\ref{m6}) at $\omega \gg 1$ can be obtained by expanding
the function under the exponential in $1/\omega $. Keeping the leading terms
we have%
\begin{equation*}
\phi (x)\approx 3+2\ln \left( \frac{x}{2\omega }\right) +\frac{2x}{\omega }.
\end{equation*}%
In the limit $\omega \gg 1$ Eq. (\ref{m6}) becomes%
\begin{equation*}
P_{exc}=\frac{g^{2}}{\omega ^{2}}\left\vert \int\limits_{0}^{\infty
}dxe^{-2i\nu \ln x}e^{-ix\left( 1+\frac{2\nu }{\omega }\right) }\right\vert
^{2}
\end{equation*}%
\begin{equation}
=\frac{g^{2}}{\omega ^{2}\left( 1+\frac{2\nu }{\omega }\right) ^{2}}%
\left\vert \int\limits_{0}^{\infty }dxx^{2i\nu }e^{ix}\right\vert ^{2}.
\end{equation}%
Using 
\begin{equation*}
\int\limits_{0}^{\infty }dxx^{2i\nu }e^{ix}=-\frac{\pi e^{-\pi \nu }}{\sinh
\left( 2\pi \nu \right) \Gamma \left( -2i\nu \right) },
\end{equation*}%
where $\Gamma (z)$ is the gamma function, and the property $|\Gamma
(-ix)|^{2}=\pi /[x\sinh (\pi x)]$ we find%
\begin{equation}
P_{exc}=\frac{4\pi g^{2}\nu }{\omega ^{2}\left( 1+\frac{2\nu }{\omega }%
\right) ^{2}}\frac{1}{e^{4\pi \nu }-1}.  \label{m7}
\end{equation}

$P_{exc}$ becomes exponentially small for $\nu \gg 1$. Thus, acceleration
radiation will not be emitted with very large $\nu $. On the other hand,
typical atomic frequencies $\omega \gg 1$ and, therefore, in the following
one can assume that $\omega \gg \nu $. Then, in the dimensional units Eq. (%
\ref{m7}) reads%
\begin{equation}
P_{exc}=\frac{4\pi g^{2}r_{g}\nu }{c\omega ^{2}}\frac{1}{e^{\frac{4\pi
r_{g}\nu }{c}}-1}.
\end{equation}%
The probability of photon absorption is obtained by changing $\nu
\rightarrow -\nu $, which for $\omega \gg \nu $ yields 
\begin{equation}
P_{abs}=e^{\frac{4\pi r_{g}\nu }{c}}P_{exc}.
\end{equation}

\section{Density matrix for the field mode}

\label{A3}

The (microscopic) change in the density matrix of a field mode $\delta \rho
^{i}$ due to an atom injected at time $\tau _{i}$ is%
\begin{equation*}
\delta \rho ^{i}=-\frac{1}{\hbar ^{2}}\int_{\tau _{i}}^{\tau _{i}+T_{\text{%
int}}}\int_{\tau _{i}}^{\tau _{i}+\tau ^{\prime }}d\tau ^{\prime }d\tau
^{\prime \prime }
\end{equation*}%
\begin{equation}
\text{Tr}_{atom}\left[ \hat{V}(\tau ^{\prime }),\left[ \hat{V}(\tau ^{\prime
\prime }),\rho ^{\text{atom}}(\tau _{i})\otimes \rho (t(\tau _{i}))\right] %
\right] ,  \label{B1}
\end{equation}%
where $T_{\text{int}}$ is the proper atom-field interaction time, Tr$_{atom}$
denotes the trace over atom states and $\hat{V}(\tau )$ is the interaction
Hamiltonian between the atom and the field mode given by Eq. (\ref{u2}). The
time $\tau $ is the atomic proper time, i.e., the time measured by an
observer riding along with the atom.

In the case of random injection times, the equation of motion for the
density matrix of the field is%
\begin{equation*}
\frac{d\rho _{n,n}}{dt}=-\Gamma _{e}\left[ (n+1)\rho _{n,n}-n\rho _{n-1,n-1}%
\right] -
\end{equation*}%
\begin{equation}
-\Gamma _{a}\left[ n\rho _{n,n}-(n+1)\rho _{n+1,n+1}\right] ,  \label{u3}
\end{equation}%
where $\Gamma _{e}$ and $\Gamma _{a}$ are emission and absorption rates of
the photon in the cavity, $\Gamma _{e,a}=r|gI_{e,a}|^{2}$, and $I_{e,a}$ are
given by the integrals 
\begin{equation*}
ge^{-i\xi /\pi }I_{e,a}=-\frac{i}{\hbar }\int_{\tau _{i}}^{\tau _{i}+T_{%
\text{int}}}V_{e,a}d\tau ,
\end{equation*}%
where $\xi =2\pi \nu r_{g}/c$ and $\nu $ is the mode frequency far from BH.
We note that the absorption and emission matrix elements of the interaction
Hamiltonian are as in Appendix \ref{A2} 
\begin{equation*}
V_{a}=\left\langle 0,a\right\vert \hat{V}(\tau )\left\vert 1,b\right\rangle
,\quad V_{e}=\left\langle 1,a\right\vert \hat{V}(\tau )\left\vert
0,b\right\rangle ,
\end{equation*}%
and obtain Eq. (\ref{u3}). Leakage of photons into \textquotedblleft outer
space\textquotedblright\ relative to the atomic cloud-BH complex as in Fig. %
\ref{Fig1} can be taken into account by adding \textquotedblleft
leakage\textquotedblright\ terms to the density matrix equation (\ref{u3}).
However, if the rate of photon loss from the cavity is much smaller then $%
\Gamma _{a}$, such terms can be omitted in Eq. (\ref{u3}).

When absorption is greater then emission there is a steady state solution of
Eq. (\ref{u3}) given by the thermal distribution \cite{Scul66} 
\begin{equation}
\rho _{n,n}^{S.S.}=\exp \left( -2\xi n\right) \left[ 1-\exp \left( -2\xi
\right) \right] .  \label{B3}
\end{equation}

In order to approach this steady state solution, we need a cavity to
restrict the modes to a finite range of the Regge-Wheeler coordinate $%
r_{\ast }$, so the bottom of the cavity must be at $r_{b}>r_{g}$, and the
top must be at $r_{t}<\infty $. This will modify the analysis of Appendix %
\ref{A2}, but we can then take the limit as $r_{b}\rightarrow r_{g}$ and $%
r_{t}\rightarrow \infty $.

\section{Entropy flux}

\label{A4}

The time rate of change of entropy inside the cavity due to photon
generation,%
\begin{equation}
\dot{S}_{p}=-k_{B}\sum_{n,\nu }\dot{\rho}_{n,n}\ln \rho _{n,n},
\end{equation}%
to a good approximation can be written as%
\begin{equation}
\dot{S}_{p}\approx -k_{B}\sum_{n,\nu }\dot{\rho}_{n,n}\ln \rho _{n,n}^{S.S.}
\label{SD}
\end{equation}%
once one has approached the steady state solution \cite{remark}. The steady
state density matrix $\rho _{n,n}^{S.S.}$ is given by Eq. (\ref{B3}).
Inserting it into (\ref{SD}) gives%
\begin{equation}
\dot{S}_{p}\approx \frac{4\pi k_{B}r_{g}}{c}\sum_{\nu }\dot{\bar{n}}_{\nu
}\nu ,
\end{equation}%
where $\dot{\bar{n}}_{\nu }$ is the photon flux from the cavity.

Recalling the BH area $A\equiv 4\pi r_{g}^{2}$, where the gravitational
radius $r_{g}=2MG/c^{2}$, and $\dot{m}_{p}c^{2}=\hbar \sum_{\nu }\dot{\bar{n}%
}_{\nu }\nu $ is the power carried away by the emitted photons, we arrive at
the HBAR entropy/area relation%
\begin{equation}
\dot{S}_{p}=\frac{k_{B}c^{3}}{4\hbar G}\dot{A}_{p}\,.
\end{equation}%
Here $\dot{A}_{p}=32\pi G^{2}M\dot{m}_{p}/c^{4}$ is the rate of change of
the BH area due to photon emission. The BH rest mass changes as $\dot{M}%
=\dot{m}_{\text{atom}}+\dot{m}_{p}$ due to the atomic cloud adding to and
the emitted photons taking from the mass of the BH. The BH area $A$ is
proportional to $M^{2}$ and, hence, $\dot{A}=(2\dot{M}/M)A=\dot{A}_{\text{%
atom}}+\dot{A}_{p}$.

\end{document}